\documentclass[aps,pre,reprint,showpacs]{revtex4-1}
\usepackage{graphicx}

\begin{document}

\title{Successful attack on permutation-parity-machine-based neural
  cryptography}

\author{Lu\'is F. Seoane}
\affiliation{Bernstein Center for Computational Neurosciences,
  Technische Universit\"at Berlin, Germany}
\author{Andreas Ruttor}
\affiliation{Artificial Intelligence Group, Technische Universit\"at
  Berlin, Germany}

\begin{abstract}
  An algorithm is presented which implements a probabilistic attack on
  the key-exchange protocol based on permutation parity machines.
  Instead of imitating the synchronization of the communicating
  partners, the strategy consists of a Monte Carlo method to sample
  the space of possible weights during inner rounds and an analytic
  approach to convey the extracted information from one outer round to
  the next one. The results show that the protocol under attack fails
  to synchronize faster than an eavesdropper using this algorithm.
\end{abstract}


\maketitle

Interacting feed-forward neural networks can synchronize by mutual
learning \cite{Metzler:2000:INN, Kinzel:2000:DIN}. If two networks A
and B are trained with examples consisting of random inputs and the
corresponding output of the other one, their weight vectors converge.
In the case of tree parity machines (TPMs) this mutual synchronization
of A and B requires fewer examples than training a third network E
successfully \cite{Rosen-Zvi:2002:CBN, Rosen-Zvi:2002:MLT,
  Kinzel:2003:DGI, Kanter:2005:TNN, Ruttor:2007:DNC}. Based on this
effect a TPM-based neural key-exchange protocol has been developed
\cite{Kanter:2002:SEI, Mislovaty:2003:PCC, Ruttor:2004:NCF,
  Ruttor:2005:NCQ} and shown to be useful in embedded devices
\cite{Volkmer:2005:TPM, Muehlbach:2008:SCM} as well as being
sufficiently secure against several attacks \cite{Ruttor:2007:DNC}.

Recently, a variant of neural cryptography has been presented in
Ref.~\cite{Reyes:2010:PPM} which uses permutation parity machines
(PPMs) \cite{Reyes:2009:PPM} instead of TPMs. This change increases
the robustness of the key-exchange protocol against the attacks which
have been tried on the TPM-based algorithm before
\cite{Mislovaty:2002:SKE, Shacham:2004:CAN, Ruttor:2006:GAN}. However,
it also reduces the number of possible values per weight from $2 L + 1
\geq 3$ to $2$, so that other attacks become more feasible. This is
especially true for the \emph{probabilistic attack}, which has been
suggested by Ref.~\cite{Klimov:2003:ANC}, but not implemented up to
now. We have used this idea and developed an attack method especially
suited for PPM-based neural cryptography. In this Rapid Communication,
we describe our attack and present results indicating its success.

A PPM is a neural network consisting of two layers: There are $K$
hidden units in the first layer, each of which has an independent
receptive field of size $N$, and only one neuron in the second layer.
Its $K N$ inputs $x_{i,j}$ with indices $i = 1, \dots, N$ and $j = 1,
\dots, K$ are binary: $x_{i,j} \in \{ 0, 1 \}$. In order to simplify
the notation they are combined into input vectors $\mathbf{x}_j =
(x_{1,j}, \dots, x_{N,j})^\top$ or the input matrix $X =
(\mathbf{x}_1, \dots, \mathbf{x}_K)$ where appropriate.

The weights $w_{i,j}$ are selected elements from the state vector
$\mathbf{s}$ of the PPM, which consists of $G \gg K N$ elements $s_i
\in \{ 0, 1 \}$. For that purpose a matrix $\pi$ of size $N \times K$
containing numbers $\pi_{i,j} \in \{ 1, \dots, G \}$ is used, so that
$w_{i,j} = s_{\pi_{i,j}}$. The weight vector $\mathbf{w}_j = (w_{1,j},
\dots, w_{N,j})^\top$ then determines the mapping from the input
vector $\mathbf{x}_j$ to the state $\sigma_j \in \{ 0, 1 \}$ of the
$j$-th hidden unit. First, the \emph{vector local field}
$\mathbf{h}_j$ is calculated as the one-by-one logical \textsc{xor}
operation
\begin{equation}
  \mathbf{h}_j = \mathbf{x}_j \oplus \mathbf{w}_j \quad
  \Longrightarrow \quad h_{i,j} = x_{i,j} \oplus w_{i,j}
\end{equation}
of the bits in $\mathbf{x}_j$ and $\mathbf{w}_j$. Then the unit
becomes active, $\sigma_j = 1$, if the majority of elements in
$\mathbf{h}_j$ is equal to $1$, otherwise it stays inactive, $\sigma_j
= 0$:
\begin{equation}
  \sigma_j = \Theta \left( h_j - \frac{N}{2} \right),
\end{equation}
where
\begin{equation}
  h_j = \sum_{i=1}^N h_{i,j}
\end{equation}
denotes the \emph{scalar local field} and
\begin{equation}
  \Theta(x) = \left\{
    \begin{array}{ccl}
      0 & \mbox{for} & x \leq 0, \\
      1 & \mbox{for} & x > 0,
    \end{array}
    \right.
\end{equation}
is the Heaviside step function. Finally, the total output of the PPM
is calculated as the binary state $\tau \in \{ 0, 1 \}$ of the single
unit in the second layer which is set to the parity
\begin{equation}
  \tau = \bigoplus_{j=1}^K \sigma_j
\end{equation}
of the hidden states $\sigma_j$.

When implementing the synchronization task, two PPMs (A and B)
designed with the same settings (i.e., with same $N$, $K$, and $G$)
are provided. The synchronization will succeed after several
\emph{inner} and \emph{outer rounds}, which are described below.

For each inner round, the elements of the matrix $\pi$ and the input
vectors $\mathbf{x}_j$ are drawn randomly and independently from their
corresponding value set. These quantities are provided publicly to all
PPMs, which includes even an attacker E. Then both A and B compute
their outputs $\tau^\mathrm{A}$ and $\tau^\mathrm{B}$ and if they
agree ($\tau^\mathrm{A} = \tau^\mathrm{B}$), they store the state
$\sigma_1^\mathrm{A}$ and $\sigma_1^\mathrm{B}$ of their first hidden
units in a buffer, which remains private for each PPM.

Thus, there are as follows: public and common input vectors
$\mathbf{x}_j$ and the $\pi$ matrix; public, but not necessarily equal
outcomes $\tau^\mathrm{A}$ and $\tau^\mathrm{B}$; private, not
necessarily equal state vectors $\mathbf{s}^\mathrm{A}$ and
$\mathbf{s}^\mathrm{B}$; and private, not necessarily equal states of
the hidden units $\sigma_j^\mathrm{A}$ and $\sigma_j^\mathrm{B}$.

The inner rounds are repeated until the buffers where $\sigma_1^A$ and
$\sigma_1^B$ are stored reach size $G$. Then, an outer round is
completed and each buffer becomes the new state vector in the
corresponding machine, substituting the old one. The dynamics of the
PPMs are such that after each outer round the state vectors
$\mathbf{s}^\mathrm{A}$, $\mathbf{s}^\mathrm{B}$ tend to be more
alike, eventually reaching full synchronization $\mathbf{s}^\mathrm{A}
= \mathbf{s}^\mathrm{B}$. The synchronization time $t_\mathrm{s}$
measured in the number of outer rounds is a random variable, as it
depends on randomly chosen initial conditions and inputs. However, its
mean value rises in a polynomial fashion with increasing size $N$ of
the input vector as well as growing size $G$ of the state vector
$\mathbf{s}$ \cite{Reyes:2009:PPM}.

Reported previous attacks on PPMs tried to mimic the behavior of the
synchronizing networks by using a single machine or an ensemble
\cite{Reyes:2010:PPM}. They showed poor performance in guessing
$\mathbf{s}^\mathrm{A}$ correctly. Namely, for the attacks on PPMs
with $K=2$ and $G=128$ analyzed in Ref.~\cite{Reyes:2010:PPM}, the
probability of success did not exceed $10^{-5}$.

In the following, we present the description of a different attack
strategy. It does not pursue to mimic the synchronizing process, but
to first guess the state vector of A (or B) during an outer round and
consecutively to reproduce A's (or B's) behavior during the given
round so that a fair guessing of the bits stored in the buffer and the
subsequent $\mathbf{s}^\mathrm{A}$ for the next outer round can be
done.

Some notation is introduced. The synchronizing parties A and B are
eavesdropped by a third agent E, which implements its own PPM with
state vector $\mathbf{s}^\mathrm{E}$ and output $\tau^\mathrm{E}$.
Additionally, the attacker uses a \emph{probabilistic state vector}
$\mathbf{p}^\mathrm{E} = (p_1, \dots, p_G)^\top$ to describe its
knowledge about A's state vector $\mathbf{s}^\mathrm{A}$. Each element
$p_i$ is an approximation of the marginal probability
$P(s_i^\mathrm{A} = 0 | D)$ that the $i$-th bit of
$\mathbf{s}^\mathrm{A}$ is $0$ given all data $D$ observed by E
before, i.e., inputs and outputs of A and B, which have already been
transmitted over the public channel.

At the beginning of the probabilistic attack previous information
about $\mathbf{s}^\mathrm{A}$ is not available. Therefore, E starts
with a neutral hypothesis and all $p_i$ are initialized with the prior
probability $P(s_i = 0) = 1/2$.

In each inner round an input $X$ and a matrix $\pi$ are provided to
all PPMs. Then A and B calculate their outputs and communicate them
publicly. This enables E to update $\mathbf{p}^\mathrm{E}$ based on
the observed data $X$, $\pi$, and $\tau^\mathrm{A}$. For that purpose
the posterior probability $P(s_i = 0 | \mathbf{p}^\mathrm{E}, X, \pi,
\tau^\mathrm{A})$ is estimated using a Monte Carlo approach, which is
similar to approximate Bayesian computation \cite{Toni:2009:ABC}.

This works by generating $M$ state vectors $\mathbf{s}^\mathrm{E}$
which are compatible with the current observation as well as prior
knowledge obtained in previous rounds. The $G$ elements of a candidate
$\mathbf{s}^\mathrm{E}$ are sampled independently from the Bernoulli
distribution with probabilities $P(s_i = 0) = p_i$ and $P(s_i = 1) = 1
- p_i$. Of course, it is only necessary to draw bits $s_i$ which are
selected as weights by $\pi$. All others can be omitted without
affecting the result. This shortcut speeds the sampling up
considerably if $G \gg K N$. Then $\mathbf{s}^\mathrm{E}$ is plugged
into E's PPM together with $X$ and $\pi$ in order to calculate
$\tau^\mathrm{E}$. If E's output matches A's, $\tau^\mathrm{E} =
\tau^\mathrm{A}$, the candidate is stored; if not, it is dismissed.
This procedure goes on until $M$ valid state vectors
$\mathbf{s}^\mathrm{E}$ have been produced.

Afterward, the desired marginal posterior probability $P(s_i = 0 |
\mathbf{p}^\mathrm{E}, X, \pi, \tau^\mathrm{A})$ can be estimated as
the relative frequency of $s_i = 0$ in the sample. The result is then
used to update all $p_i$ which have been selected as weights in the
current round. The other elements of $\mathbf{p}^\mathrm{E}$ remain
unchanged, because the attacker gained no information about the
corresponding parts of $\mathbf{s}^\mathrm{A}$. Of course, this
computation is repeated for the next inner round.

As the space of all weight matrices $W = (\mathbf{w}_1, \dots,
\mathbf{w}_K)$ is of size $2^{N K}$, approximately $2^{N K - 1}$ of
them are compatible with a given $X$, $\pi$, and $\tau^\mathrm{A}$.
Thus if the sampling algorithm generates $M \geq 2^{N K - 1}$ state
vectors, it would be similar to a brute force attack. But choosing
such a large parameter $M$ is only feasible for a very small number of
weights.

Updating $\mathbf{p}^\mathrm{E}$ as described above has the effect
that its elements $p_i$ converge toward $0$ or $1$ after several
rounds, so that finally $M$ equal state vectors with
\begin{eqnarray}
  p_i = 0 &\quad\Longrightarrow\quad& s_i = 1, \\
  p_i = 1 &\quad\Longrightarrow\quad& s_i = 0,
\end{eqnarray}
are sampled. However, defining
\begin{equation}
  s_i^\mathrm{E*} = \left\{
    \begin{array}{ccl}
      0 & \mbox{for} & p_i   >  1/2, \\
      1 & \mbox{for} & p_i \leq 1/2,
    \end{array}
  \right.
\end{equation}
as the most probable state provided $\mathbf{p}^\mathrm{E}$, the
attack is considered a success as soon as $\mathbf{s}^\mathrm{E*} =
\mathbf{s}^\mathrm{A}$ without regard to whether or not all the $p_i$
have collapsed to $0$ or $1$.

In contrast, if one or more $p_i$ have collapsed to the wrong value, E
might be unable to achieve the desired output $\tau^\mathrm{E} =
\tau^\mathrm{A}$ in a later round. Such a failure clearly indicates
that the estimation of some $p_i$ has gone wrong. In order to avoid an
infinite loop in this case, only a finite number of attempts is made
to generate $M$ valid samples $\mathbf{s}^\mathrm{E}$. If the limit is
reached, the element $p_i$ of $\mathbf{p}^\mathrm{E}$ which is closest
to collapse is reset to the neutral hypothesis, $p_i = 1/2$.

Usually, the algorithm will not be able to guess
$\mathbf{s}^\mathrm{A}$ correctly in less than one outer round,
therefore we need a mechanism to transfer the information gained
during an outer round into the next one. Let $\mathbf{p}^\mathrm{E-}$
be the probabilistic state vector after applying the previous
algorithm on all the inner rounds of a whole outer round. In order to
transfer the information the attacker calculates the probability
distribution for the state $\sigma_1^\mathrm{E}$ of the first hidden
unit conditioned on the probabilistic state vector
$\mathbf{p}^\mathrm{E-}$ as well as the input $X$ and the matrix
$\mathbf{\pi}$ for each of the inner rounds with $\tau^\mathrm{A} =
\tau^\mathrm{B}$. The result is then used to construct the
probabilistic state vector $\mathbf{p}^\mathrm{E+}$ for the start of
the next outer round.

\begin{figure}
  \centering
  \includegraphics[width=0.45\textwidth]{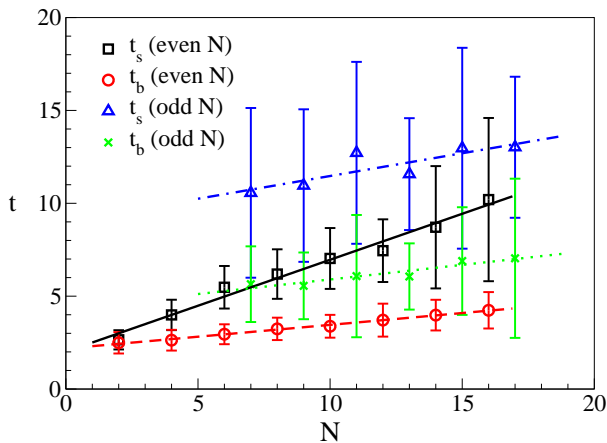}
  \caption{(Color online) Synchronization time $t_\mathrm{s}$ and
    break time $t_\mathrm{b}$ measured in outer rounds as a function
    of $N$ for PPMs with $K=2$, $G=128$. Symbols denote mean values
    and error bars denote standard deviations obtained in $100$ runs
    (even $N$). Lines show the results of a fit using linear
    regression as given in Table~\ref{tab:regression}. For odd $N$
    around 25 out of 100 runs reaching $t=30$ had to be aborted and
    discarded from the data set.}
  \label{fig:time}
\end{figure}

In the following we describe an algorithm to approximate the
probability that a single hidden unit has internal state $\sigma_j =
0$ given $\mathbf{p}^\mathrm{E}$ and the corresponding public
information of an inner round. The output $\sigma_j$ depends only on
the number of $1$s in the vector local field $\mathbf{h}_j$, which is
equal to the scalar local field $h_j$. Here we approximate the
probability distribution $P(h_j = n | \mathbf{p}^\mathrm{E-}, X, \pi)$
of this quantity by a binomial distribution which uses the average
probability of finding a $1$ in $\mathbf{h}_j$ as a parameter
\begin{equation}
  q_j = \frac{1}{N} \sum_{i=1}^N \left[ x_{i,j} \, p_{\pi_{i,j}} + (1
    - x_{i,j}) (1 - p_{\pi_{i,j}}) \right].
\end{equation}
Then, the probability
\begin{eqnarray}
  P(\sigma_j = 0 | \mathbf{p}^\mathrm{E-}, X, \pi) = \sum_{n=0}^{N /
    2} P(h_j = n | \mathbf{p}^\mathrm{E-}, X, \pi)
\end{eqnarray}
of $\sigma_j = 0$ is given by
\begin{equation}
  P(\sigma_j = 0 | \mathbf{p}^\mathrm{E-}, X, \pi) = \sum_{n=0}^{N /
    2} {N \choose n} q_j^n (1 - q_j)^{N - n}.
\end{equation}
Finally, the attacker stores this result in $\mathbf{p}^\mathrm{E+}$
whenever $\tau^\mathrm{A} = \tau^\mathrm{B}$ occurred in the inner
round. This procedure succeeded in conveying enough information from
one outer round to the next one.

An alternative approach to this task seems to be Monte Carlo sampling
of $\sigma_1^\mathrm{E}$ conditioned on the final
$\mathbf{p}^\mathrm{E-}$. But in our simulations this method proved to
be prone to failure: Either $\mathbf{p}^\mathrm{E}$ was effectively
reset or the algorithm could not generate enough valid weight
candidates at some inner round. Thus we developed and used the
analytic approach instead of calculating $\mathbf{p}^\mathrm{E+}$ by
sampling.

The attack described in this Rapid Communication is often capable of
guessing the state vector $\mathbf{s}^\mathrm{A}$ in a number of outer
rounds that are less than the number of rounds that A and B needed to
synchronize. This result was reproduced for many different setups of
the synchronizing PPMs: varying input vector size and varying state
vector size. The usual setup for cryptographic is to use even $N$,
since PPMs with odd $N$ synchronize notably slower or sometimes not at
all \cite{Reyes:2009:PPM}. However, the algorithm was also tried for
odd $N$ with an illustrative purpose and yielded satisfactory results.

\begin{figure}
  \centering
  \includegraphics[width=0.45\textwidth]{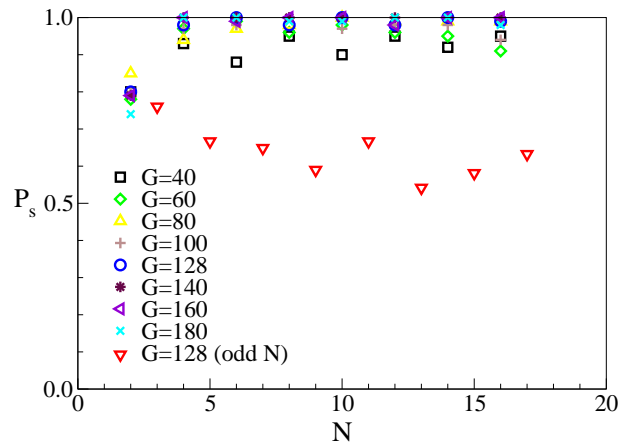}
  \caption{(Color online) Success probability $P_\mathrm{s}$ of the
    attack as a function of $N$ for PPMs with $K=2$. Symbols denote
    the percentage of successful attacks found in $100$ simulations
    (even $N$). For odd $N$ around 25 out of 100 runs had to be
    stopped after $30$ outer rounds without a clear result, i.e.,
    $t_\mathrm{s} > 30$ and $t_\mathrm{b} > 30$. These simulations
    were not considered for calculating the probability of success
    $P_\mathrm{s}$.}
  \label{fig:success}
\end{figure}

\begin{table}
  \centering
  \begin{tabular}{ccl@{$\,\pm\,$}ll@{$\,\pm\,$}l}
    \hline
    $N$ & Time & \multicolumn{2}{c}{Slope $a$} &
    \multicolumn{2}{c}{Offset $b$} \\
    \hline
    Even & $t_\mathrm{s}$ & $0.495$  & $0.028$  & $2.01$  & $0.28$  \\
    Even & $t_\mathrm{b}$ & $0.1275$ & $0.0038$ & $2.180$ & $0.038$ \\
    Odd  & $t_\mathrm{s}$ & $0.245$  & $0.076$  & $9.02$  & $0.95$  \\
    Odd  & $t_\mathrm{b}$ & $0.157$  & $0.028$  & $4.33$  & $0.35$  \\
    \hline
  \end{tabular}
  \caption{Linear regression with model $t = a N + b$ for average
    synchronization time $t_\mathrm{s}$ and break time
    $t_\mathrm{b}$.}
  \label{tab:regression}
\end{table}

As for the technicalities a sampling size of $M = 10^3$ was chosen.
This implies that for $N = 2, 4$ the algorithm works similar to a
brute force attack, but for large $N$ only a small part of the weight
space is sampled, e.g., for $N = 8$ only around $3\%$ of all possible
weight configurations. The absent of performance drop notwithstanding
the scarce sampling highlights the efficiency of the algorithm. The
mechanism to prevent the attack from getting stuck was implemented by
resetting one of the bits each time that $M^2 = 10^6$ consecutive
unsuccessful attempts at generating a valid weight candidate were
reached. Finally, the attack was considered a success as soon as
$\mathbf{s}^\mathrm{E*} = \mathbf{s}^\mathrm{A}$ has been reached. The
number of outer rounds needed to achieve this is called break time
$t_\mathrm{b}$, which varies randomly depending on the initial
conditions and the course of the key exchange.

Figure \ref{fig:time} shows that the mean values of synchronization
time $\langle t_\mathrm{s} \rangle$ as well as break time $\langle
t_\mathrm{b} \rangle$ grow linearly with increasing size $N$ of the
input vectors. For all cases presented here we find that the attacker
is faster than the two partners on average, $\langle t_\mathrm{b}
\rangle < \langle t_\mathrm{s} \rangle$. Additionally, linear
regression results shown in Table~\ref{tab:regression} indicate that
$\langle t_\mathrm{b} \rangle$ grows slower than $\langle t_\mathrm{s}
\rangle$, so that increasing $N$ does not improve the security of the
PPM-based key exchange.

Synchronization with odd $N$ is much slower than for even $N$. Only
runs with $t_\mathrm{b} < 30$ or $t_\mathrm{s} < 30$ were considered
here to reduce computational costs. This condition also excludes
failed synchronization attempts caused by reaching a stable
\emph{antiparallel} weight configuration \cite{Reyes:2009:PPM}, which
can only happen if $N$ is odd.

In Fig.~\ref{fig:success} the performance of the algorithm is
presented in terms of the probability of success of the attack. The
functionality for many more different setups is examined here.
Regarding cases with even $N$, the performance of the algorithm
generally increases as $N$ or $G$ become larger. For nearly all
configurations shown here the success probability $P_\mathrm{s}$ is
above $80\%$ and it actually reaches $100\%$ in many situations. Odd
$N$ is considerably more difficult for the attacker, but nevertheless
the success probability $P_\mathrm{s}$ remains larger than $50\%$.
These values, however, have been obtained for single runs of our
algorithm on each data set. As the method is non-deterministic due to
Monte Carlo sampling in each inner round, repeating it on the same
observations should lead to even more success.

Consequently, the results clearly show that the PPM-based neural
key-exchange protocol using the parameter values $K$, $N$, and $G$
analyzed in this Rapid Communication is not secure enough for any
cryptographic application. Furthermore, there is no indication that
increasing the sizes of input or state vectors would reduce the
success probability and lead to a secure configuration.

In contrast, the complexity of successful attacks on TPM-based neural
cryptography increases exponentially with the number $2 L + 1$ of
possible weight values, but the effort of the partners grows only
proportional to $L^2$ \cite{Ruttor:2007:DNC}. Here $L$ has the same
effect as the key size in encryption algorithms, which allows to
balance speed and security. While the probabilistic attack
\cite{Klimov:2003:ANC} has not been tested on TPM-based neural
cryptography, it is quite likely that the same scaling law for $L$
applies to its success probability. But in order to answer this open
question we are going to implement and analyze such probabilistic
attacks also for TPMs.

The same question could be asked regarding the security of \emph{chaos
  cryptography} \cite{Cuomo:1993:CIS, Grassi:1999:STA,
  Klein:2006:PCM}, which is based on a similar synchronization
principle \cite{Pecora:1990:SCS}. Consequently, probabilistic attacks
should be envisioned and tested there, too. Nevertheless, the
specificity of the present implementation suggests that further
development is needed for attacks on chaotic cryptography.

L.F.S.~acknowledges the financial support of Fundaci\'on Pedro
Barri\'e de la Maza and funding Grant No.~01GQ1001B.

\bibliographystyle{apsrev4-1}
\bibliography{paper}
	
\end{document}